\documentclass{iopbk2e}
\usepackage{iopams}
\usepackage[dvips]{graphicx,psfrag}
\usepackage{epsf}
\usepackage[mathcal,mathscr]{eucal}
\usepackage{amsmath}

\newcommand{\be}{\begin{equation}}
\newcommand{\ee}{\end{equation}}
\newcommand{\beann}{\begin{eqnarray*}}
\newcommand{\eeann}{\end{eqnarray*}}
\newcommand{\bea}{\begin{eqnarray}}
\newcommand{\eea}{\end{eqnarray}}

\title{Quantum aspects of black holes\\ Claus Kiefer}
\author{To appear in: The galactic black hole,\\ 
  edited by Heino Falcke and Friedrich W Hehl\\
        (IOP Publishing, Bristol, 2002)}

\begin{document}

\maketitle

\pagenumbering{roman}
\setcounter{page}{5}
\tableofcontents

\newpage

\pagenumbering{arabic}
\setcounter{page}{1}

\Chapter[Quantum aspects]{Quantum aspects of black holes}{
Claus Kiefer\\ University of Cologne}

\section{Introduction}
\label{intro}

At the most fundamental level, black holes are genuine quantum objects.
This holds irrespective of the fact that direct quantum effects
can only be observed for small black holes -- black holes that cannot
be formed by stellar collapse. For this reason the discussion in this
chapter will be of more theoretical nature. But even a black hole as
gigantic as the galactic black hole will in the far future (if the
Universe will not recollapse) be dominated by quantum effects and
eventually evaporate. It is, however, possible that small black
holes have been created in the very early Universe. For such primordial
black holes quantum effects can be of direct observational significance
in the present Universe. I shall thus devote my last section to a brief
discussion of their relevance.
In the first three sections I shall, however, give an introduction
to the key theoretical developments -- black-hole mechanics,
Hawking radiation, and the interpretation of the black-hole entropy.

In my discussion I shall draw heavily from my review article
Kiefer (1999) where many technical details 
can be found. Other general references include the comprehensive book by
Frolov and Novikov (1998), Wald (2001), Hehl \etal (1998),
as well as the article by Bekenstein (1980) and the book by Thorne (1994).

\section{The laws of black-hole mechanics}
\label{mechanics}

It is a most amazing fact that black holes obey {\em uniqueness theorems} 
(Heusler 1996). If an object collapses to form a black hole, a stationary
state is reached asymptotically. One can prove within the Einstein-Maxwell
theory that stationary black holes are uniquely characterised by
only three parameters: Mass $M$, angular momentum $J\equiv Ma$,
 and electric charge
$q$. In this sense, black holes are objects much simpler than ordinary
stars -- given these parameters, they all look the same. All other degrees
of freedom that might have been initially present have thus been radiated
away during the collapse, e.g. in the form of electromagnetic or
gravitational radiation. Since the latter constitute some form of
``hair'', one refers to the content
of these theorems as {\em black holes have
no hair}. The three parameters are associated with conservation laws
at spatial infinity. In principle, one can thus decide about the
nature of a black hole far away from the hole itself, without having
 to approach it.
 In astrophysical situations, the two parameters $M$ and $J$
suffice, since a charged object would rapidly discharge.
The corresponding solution of Einstein's equations is called the
Kerr solution (Kerr-Newman in the presence of charge).
Stationary black holes are axially symmetric with spherical symmetry 
being obtained as a special case for $J=0$.

In the presence of other fields, the uniqueness theorems do not
always hold, see e.g. N\'{u}\~{n}ez \etal (1998).
 This is in particular the case in the presence of nonabelian gauge fields.
In addition to charges at infinity, such ``coloured black holes''
have to be characterised by additional variables, and it is necessary
to approach the hole to determine them. The physical reason for the
occurrence of such solutions is the nonlinear character of these gauge
fields. Fields in regions closer to the black hole (that would otherwise
be swallowed by the hole) are tied to fields far away from the
hole (that would otherwise be radiated away) to reach an equilibrium
situation. In most examples this equilibrium is, however, 
unstable and the corresponding black-hole solution
does not represent a physical solution. Since classical nonabelian
fields have never been observed (the description of objects such as
quarks necessarily needs quantised gauge fields which, due to
confinement, have no macroscopic limits), they will not be taken
into account in the following discussion.  

In 1971, Stephen Hawking could prove an important theorem about
stationary black holes -- that their area can never decrease with time.
More precisely, he showed that

\begin{quote}
For a predictable black hole satisfying $R_{ab}k^ak^b \ge 0$ for all null
$k^a$, the surface area of the \textit{future} event horizon \textit{never}
decreases with time.
\end{quote}

A `predictable' black hole is one for which the cosmic censorship
hypothesis holds -- this is thus a major assumption for
the area law. Cosmic censorship states that all black holes
occurring in nature have an event horizon, so that
the singularity cannot be observed for far-away observers (the
singularity is not ``naked'').
I  emphasise that the time asymmetry in this theorem
comes into play because
a statement is made about the future horizon, not the past horizon; the
analogous statement for white holes would then be that the past event
horizon never increases. I also emphasise that the area law only holds in
the classical theory, not in the quantum theory (see section
\sref{hawking}).

The area law seems to exhibit a close formal analogy to the Second Law of
thermodynamics -- there the \textit{entropy} can \textit{never} decrease
with time (for a closed system). However, the conceptual difference could
not be more pronounced: while the Second Law is related to statistical
behaviour, the area law is just a theorem in differential geometry. That
the area law is in fact directly related to the Second Law will become
clear in the course of this section.

Further support for this analogy is given by the existence of analogies to
the other laws of thermodynamics. The Zeroth Law states that there is a
quantity, the temperature, that is constant on a body in thermal
equilibrium. Does there exist an analogous quantity for a black hole?
One can in fact prove that the surface gravity $\kappa$
is constant over the event horizon (Wald~1984).
 For a Kerr black hole, $\kappa$ is given by
\begin{equation}\label{kappa-explicit}
  \kappa = \frac{\sqrt{(GM)^2-a^2}}{2 G M r_+}\quad\stackrel{a \rightarrow 0}
  { \longrightarrow}\quad \frac{1}{4 G M} = \frac{GM}{R_0^2}\, ,
\end{equation}
where $r_+$ denotes the location of the event horizon.
One recognises in the Schwarzschild limit the well-known expression for
the Newtonian gravitational acceleration. ($R_0\equiv 2GM$ there
denotes the Schwarzschild radius).
One can show for a static black hole
 that $\kappa$ is the limiting force that must be exerted at infinity
to hold a unit test mass in place when approaching the horizon. This
justifies the name surface gravity.

With a tentative formal relation between surface gravity and temperature,
and between area and entropy, the question arises whether a First Law of
thermodynamics can be proved. This can in fact be done and the result for a
Kerr-Newman black hole is
\begin{equation}\label{first-law}
  \rmd M =\frac{\kappa}{8\pi G} \rmd A + \Omega_\mathrm{H}\rmd J+ \Phi
  \rmd q\,,
\end{equation}
where $A,\Omega_H,\Phi$ denote the area of the event horizon, the
angular velocity of the black hole, and
 the electrostatic potential, respectively.
This relation can be obtained by conceptually different
methods: A
\textit{physical process version} whereby a stationary black hole is
altered by infinitesimal physical processes, and an
\textit{equilibrium state version} whereby the areas of two
stationary black-hole solutions to Einstein's
equations are compared. Both methods lead to the same result
\eref{first-law}.

Since $M$ is the energy of the black hole, \eref{first-law} is the analogue
of the First Law of thermodynamics given by
\begin{equation}\label{td-first-law}
  \rmd E = T\rmd S - p\rmd V + \mu \rmd N\,.
\end{equation}
`Modern' derivations of \eref{first-law} make use of both Hamiltonian and
Lagrangian methods of general relativity. For example, the First
Law follows from an arbitrary diffeomorphism invariant theory of gravity
whose field equations can be derived from a Lagrangian.

What about the Third Law of thermodynamics? A `physical process version'
was proved by Israel -- it is impossible to reach $\kappa = 0$ in a finite
number of steps,
although it is unclear whether this is true under all
circumstances (Farrugia and Hajicek~1979).
 This corresponds to the `Nernst version' of the Third Law.
The stronger `Planck version', which states
that the entropy goes to zero (or a material-dependent constant) if the
temperature approaches zero, does not seem to hold.
 The above analogies are summarised in
table~\ref{analogie-tabelle}.
\begin{table}[ht]
\begin{center}
\begin{tabular}{c|c|c}
 Law & Thermodynamics & Stationary Black Holes \\[1ex]\hline
 \rule{0cm}{5.5ex}Zeroth& \parbox{3.5cm}{\centering $T$ constant on a body in thermal equilibrium}
 & \parbox{3.5cm}{\centering $\kappa$ constant on the horizon of a black hole}
 \\[3ex]
 First & $\rmd E = T\rmd S - p\rmd V + \mu \rmd N$ &
 $\displaystyle\rmd M =\frac{\kappa}{8\pi G} \rmd A + \Omega_\mathrm{H}\rmd J+ \Phi
  \rmd q$\\[2ex]
 Second & $\rmd S \ge 0 $ &$ \rmd A \ge 0 $ \\[2ex]
 Third & $ T = 0 $ cannot be reached & $\kappa = 0 $ cannot be reached
 \\[2ex]
\end{tabular}
\caption{\label{analogie-tabelle}}
\end{center}
\end{table}

The identification of the horizon area with the entropy
for a black hole can be obtained from
a conceptually different point of view. If a box with, say, thermal
radiation of entropy $S$ is thrown into the black hole, it seems as if the
Second Law could be violated, since the black hole is characterised only by
mass, angular momentum, and charge, but nothing else. The demonstration
that the Second Law is fulfilled
leads immediately to the concept of a black-hole entropy, as
will be discussed now (Bekenstein 1980; Sexl and Urbantke 1983).

\begin{figure}[t]
\begin{center}
\mbox{\epsfbox{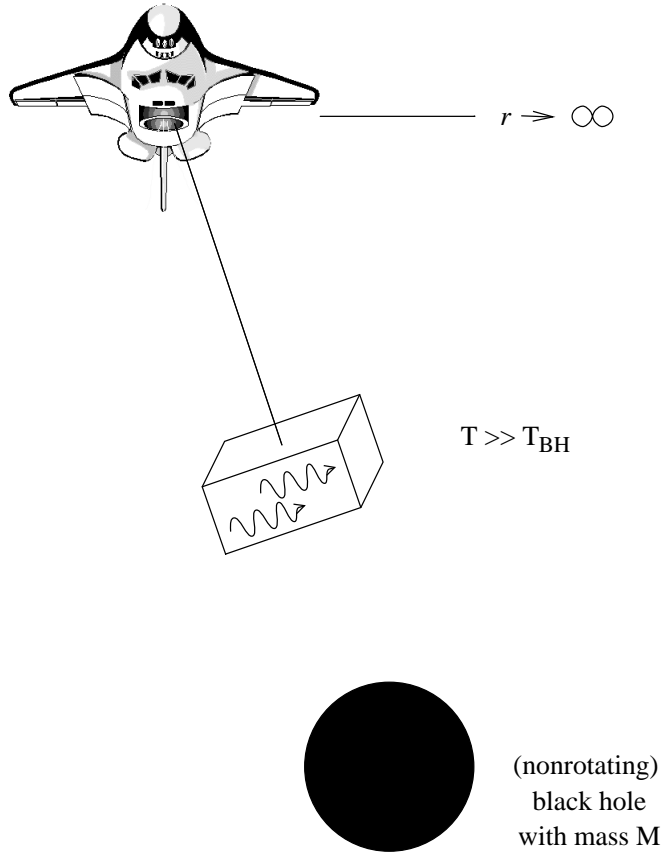}}
\caption{\label{station}Gedankenexperiment to demonstrate the Second Law
of thermodynamics for black holes}
\end{center}
\end{figure}

Consider a box with thermal radiation of mass $m$ and temperature $T$
lowered from a spaceship far away from a spherically-symmetric black hole
towards the hole (figure~\ref{station}). As an idealisation, both the rope
and the walls are assumed to have negligible mass.
At a coordinate distance $r$ from the black hole, the energy of the box is
given by
\begin{equation}\label{box-energy}
  E_r = m \sqrt{1-\frac{2GM}{r}}\quad \stackrel{r \rightarrow
  R_0}{\longrightarrow} \quad 0\,.
\end{equation}
If the box is lowered down to the horizon, the energy gain is thus given by
$m$. The box is then opened and thermal radiation of mass  $\delta m$
escapes into the hole. If the box is then closed and brought back again to
the spaceship, the energy loss is $m -\delta m$. In total the energy
$\delta m$ of the thermal radiation can be transformed into work with a
degree of efficiency $\eta =1$ . This looks as if one possessed a perpetuum
mobile of the second kind.

The key to the resolution of this apparent paradox lies in the observation
that the box must be big enough to contain the wavelength of the enclosed
radiation. This, in turn, leads to a lower limit on the distance that the
box can approach the horizon. Therefore, only part of $\delta m$ can be
transformed into work, as I shall show now.

According to Wien's law, one must have a linear extension of the box of at
least
\begin{equation}\label{linear-extension}
  \lambda_\mathrm{max} \approx \frac{\hbar}{k_\mathrm{B} T}\,.
\end{equation}
I emphasise that at this stage Planck's constant $\hbar$ comes into play.
The box can then be lowered down to the coordinate distance  $\delta r$
(assumed to be $ \ll 2G M$) from the black hole, where according to the
Schwarzschild metric the relation between $\delta r$ and
$\lambda_\mathrm{max}$ is
\[
 \lambda_\mathrm{max} \approx \int\limits_{2GM}^{2GM + \delta r}
 \left(1-\frac{2GM}{r} \right)^{-\frac{1}{2}}\rmd r \; \approx 2 \sqrt{2GM
 \delta r}\quad \Longrightarrow \quad \delta r \approx
 \frac{\lambda_\mathrm{max}^2}{8 G M}\,.
\]
According to \eref{box-energy}, the energy of the box at $r = 2GM + \delta
r$ is
\[
E_{2GM + \delta r}= m \sqrt{1- \frac{2GM}{2GM + \delta r}} \approx \frac{m
\lambda_\mathrm{max}}{4GM} \approx  \frac{m \hbar}{4 G k_\mathrm{B} T M }\,.
 \]

Recalling that according to \eref{first-law}  the formal temperature of the
black hole, $T_\mathrm{BH}$, is proportional to the surface gravity $\kappa
= 1/(4 G M) $, the energy of the box before opening is
\[
E_{2GM + \delta r}^\mathrm{(before)} \approx m \frac{T_\mathrm{BH}}{T}\,,
 \]
 while after opening it is
\[
  E_{2GM + \delta r}^\mathrm{(after)}
  \approx (m - \delta m) \frac{T_\mathrm{BH}}{T}\,.
\]
The degree of efficiency of transforming thermal radiation into work is
thus given by
\[
\eta \approx \left.\left(\delta m - \delta
m\frac{T_\mathrm{BH}}{T}\right)\right/\delta m = 1 -
\frac{T_\mathrm{BH}}{T} < 1\,,
 \]
which is the well-known Carnot limit for the efficiency of heat
engines. From the First Law \eref{first-law} one then finds for the entropy
of the black hole $S_\mathrm{BH} \propto A = 16 \pi (GM)^2.$ It is this
agreement of conceptually different approaches to black-hole thermodynamics
that gives confidence into the 
physical meaning of these concepts. In the next section I
shall show how all these formal results can be physically interpreted in
the context of quantum theory.

\section{Hawking radiation}
\label{hawking}

We have already seen in the gedankenexperiment discussed in the last
section that $\hbar$ enters the scene, see \eref{linear-extension}. That
Planck's constant has to play a role, can be seen also from the First Law
\eref{first-law}. Since $T_\mathrm{BH} \rmd S_\mathrm{BH} = \kappa/(8 \pi
G)\, \rmd A$, one must have
\[ T_\mathrm{BH} = \frac{\kappa}{G \zeta}\,,\quad
S_\mathrm{BH} = \frac{\zeta A}{8\pi} \] with an undetermined factor
$\zeta$. What is the dimension of $\zeta$? Since $S_\mathrm{BH}$ has the
dimension of Boltzmann's constant $k_\mathrm{B}$, $k_\mathrm{B}/\zeta$ must
have the dimension of a length squared. There is, however, only one
fundamental length available, the Planck length
\begin{equation}\label{planck-length}
  l_\mathrm{p}= \sqrt{G\hbar}\quad \approx \quad 10^{-33}\, \mathrm{cm}.
\end{equation}
(For string theory, this may be replaced by the
fundamental string length.)
Therefore,
\begin{equation}\label{ts-proportionalto}
  T_\mathrm{BH}\propto \frac{\hbar \kappa}{k_\mathrm{B}}\,,\quad
S_\mathrm{BH} \propto \frac{k_\mathrm{B} A}{G \hbar}\,.
\end{equation}
The determination of the precise factors in \eref{ts-proportionalto} 
was achieved in the pioneering paper by Hawking (1975).
The key ingredient in his discussion
 is the behaviour of {\em quantum} fields on
the background of an object collapsing to form a black hole.
Similar to the situation of an external electric field (Schwinger
effect), there is no uniquely defined notion of {\em vacuum}.
This leads to the occurrence of particle creation. The peculiarity
of the black-hole case is the {\em thermal} distribution of the created
particles.

\medskip
There exists an 
analogous  effect already in flat spacetime, discussed by Unruh (1976),
following work by Fulling (1973) and Davies (1975). In the following
I shall briefly describe this effect.

\begin{figure}[ht]
\begin{center}
\mbox{\epsfbox{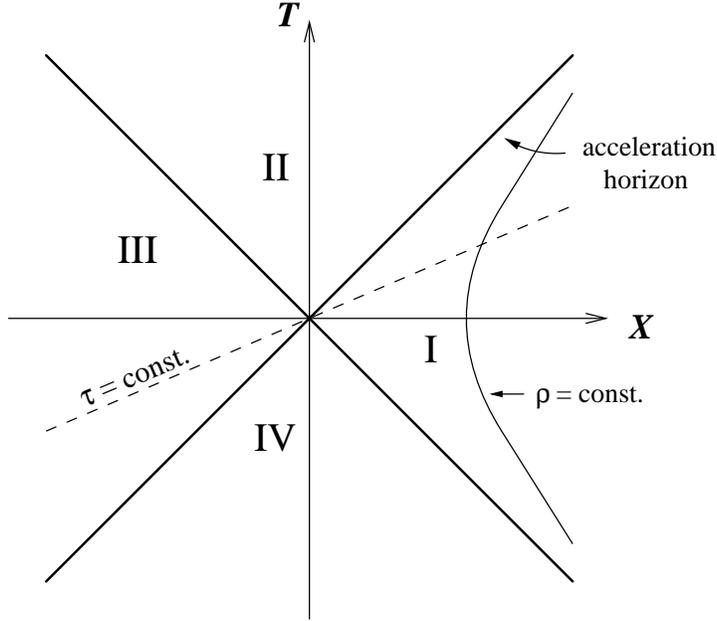}}
\caption{\label{rindler}Uniformly accelerated observer in Minkowski space}
\end{center}
\end{figure}

Whereas all inertial
observers in Minkowski space agree on the notion of vacuum
(and therefore on particles), this no longer holds for 
\textit{non-inertial} observers. 
Consider an observer who is uniformly accelerating along the
$X$-di\-rec\-tion in (1+1)-dimensional Minkowski spacetime (figure
\ref{rindler}). 
The Minkowski cartesian coordinates are labelled here by
upper-case letters. The orbit of this observer is the hyperbola shown in
figure \ref{rindler}. One recognises that, as in the Kruskal 
diagram for the Schwarzschild metric, the
observer encounters a horizon
(here an ``acceleration horizon''). There is, however, no singularity behind
this horizon. The region I is a globally hyperbolic spacetime on its own --
the so-called
\textit{Rindler spacetime}. This spacetime can be described by coordinates
$(\tau,\rho)$ which are connected to the cartesian coordinates via the
coordinate transformation
\begin{equation}\label{coordinate-transformation}
\begin{pmatrix}T\\X\end{pmatrix} = \rho \begin{pmatrix}\sinh a\tau\\ \cosh a \tau
\end{pmatrix}\,,
\end{equation}
where $a$ is a constant (the orbit in figure \ref{rindler} describes an
observer with acceleration $a$, who has $\rho = 1/a$).

Since
\begin{equation}\label{mink-metric}
  \rmd s^2 = \rmd T^2 -\rmd X^2 = a^2 \rho^2 \rmd \tau^2 - \rmd \rho^2\,,
\end{equation}
the orbits $\rho = \mathrm{constant}$ are also orbits of a timelike Killing
field $\partial/\partial \tau$. It is clear that $\tau$ corresponds to the
external Schwarzschild coordinate $t$ and that $\rho$ corresponds to $r$.
As in the Kruskal case, $\partial/\partial\tau$
becomes spacelike in regions II and IV.

\begin{sloppypar}
The analogy with Kruskal becomes even more transparent if the Schwarzschild
metric is expanded around the horizon at $ r = 2GM$.
Introducing $\rho^2/(8GM) = r -2 GM$  and recalling
\eref{kappa-explicit}, one has
\begin{equation}\label{ss-metric-expanded}
  \rmd s^2 \approx \kappa^2 \rho^2 \rmd t ^2 -\rmd \rho^2 - \frac{1}{4
  \kappa^2} \rmd \Omega^2\,.
\end{equation}
Comparison with (\ref{mink-metric}) shows that the first two terms on the
right-hand side of (\ref{ss-metric-expanded}) correspond exactly to the
Rindler spacetime (\ref{mink-metric}) with the acceleration $a$ replaced by
the surface gravity $\kappa$. The last term in (\ref{ss-metric-expanded})
describes a two-sphere with radius $(2\kappa)^{-1}$.\footnote{It is this
term that is responsible for the non-vanishing curvature of
(\ref{ss-metric-expanded}) compared to the flat-space metric
(\ref{mink-metric})  whose extension into the (neglected) other dimensions
would be just $-\rmd Y^2 -\rmd Z^2$.}
\end{sloppypar}

How does the accelerating observer experience the standard
Minkowski vacuum $|0\rangle_M$? The key point is that the vacuum
is a {\em global} state correlating regions I and III in
figure \ref{rindler} (similar to Einstein-Podolsky-Rosen correlations),
but that the accelerated observer is restricted to region~I.
Considering for simplicity the case of a massless scalar
field,
the global vacuum state comprising the regions I and II
can be written in the form
\begin{equation}\label{global-vacuum}
  |0\rangle_M = \prod_\omega \sqrt{1 - \rme^{-2\pi \omega a^{-1}}}
\sum_n \rme^{- n \pi \omega a^{-1}} |n^\mathrm{I}_\omega\rangle \otimes
|n^\mathrm{II}_\omega\rangle \,,
\end{equation}
where $|n^\mathrm{I}_\omega\rangle$ and $|n^\mathrm{II}_\omega\rangle$ are
$n$-particle states with frequency
 $\omega=|{\mathbf k}|$ in regions I and II,
respectively. The expression (\ref{global-vacuum}) is an example for
 the Schmidt
expansion of two entangled quantum systems, see e.g.\ Giulini \etal
(1996); note also the analogy of (\ref{global-vacuum}) with a BCS-state in the
theory of superconductivity.

For an observer restricted to region~I, 
the state (\ref{global-vacuum})
 cannot be distinguished, by
operators with support in I only, from a density matrix that is found from
(\ref{global-vacuum}) by tracing out all degrees of freedom in region II,
\begin{equation}\label{I-density-matrix}
 \begin{split}
   \rho_\mathrm{I}& \equiv \Tr_\mathrm{II} |0\rangle_M\langle 0|_M \\
   & = \prod_\omega \left(1 - \rme^{-2\pi \omega a^{-1}} \right)
\sum_n \rme^{-2 \pi n \omega a^{-1}} |n^\mathrm{I}_\omega\rangle
 \langle n^\mathrm{I}_\omega|\,.
 \end{split}
\end{equation}
Note that the density matrix $\rho_\mathrm{I}$ has exactly the form 
corresponding to a thermal canonical ensemble with temperature
\begin{equation}\label{rindler-temperature}
  T_\mathrm{U}=\frac{\hbar a}{2 \pi k_\mathrm{B}} \quad \approx
  4 \times 10 ^{-23} a\left[\frac{\mathrm{cm}}{\mathrm{s}^2}\right]
   {\mathrm K}\,.
\end{equation}
An observer who is accelerating uniformly through Minkowski space thus
sees a \textit{thermal} distribution of particles. This is an important
manifestation of the non-uniqueness of the vacuum state in quantum field
theory, even for flat spacetime. A more detailed discussion invoking
models of particle detectors confirms this result.

The ``Unruh temperature'' (\ref{rindler-temperature}), although
being very small for most accelerations, might be observable for
electrons in storage rings where spin precession is used as `detector'
(Leinaas~2001).
Due to the circular nature of the accelerator, the spectrum of the
observed particles is then, however, not thermal. Since this would 
complicate the direct comparison with the Hawking effect,
there exist other proposals to measure \eref{rindler-temperature},
for example with ultraintense lasers (Chen and Tajima~1999).

\medskip
I shall now turn to the case of black holes. From the form of the line
element near the horizon, (\ref{ss-metric-expanded}), one can already
anticipate that -- according to the equivalence principle --  a
black hole radiates with temperature (\ref{rindler-temperature}) in which
$a$ is replaced by $\kappa$. This is in fact what Hawking (1975)
found. The temperature reads
\begin{equation}\label{hawking-temperature}
  T_\mathrm{BH} = \frac{\hbar\kappa}{2 \pi k_\mathrm{B}}\,.
\end{equation}
For the total luminosity of the black hole one finds
\begin{equation}\label{bh-luminosity}
  L = -\frac{\rmd M}{\rmd t} = \frac{1}{2 \pi} \sum_{l=0}^\infty (2l+1)
  \int\limits_0^\infty \rmd \omega\, \omega \frac{\Gamma_{\omega l}}
  {e^{2\omega \pi\kappa^{-1}}-1}\,.
\end{equation}
The term $\Gamma_{\omega l}$ -- called `greybody factor' because it encodes
a deviation from the black-body spectrum -- takes into account
the fact that some of the particle modes are back-scattered into the
black hole through the effect of spacetime curvature.

\medskip
For the special case of the Schwarzschild metric where $\kappa = (4GM)^{-1}$,
(\ref{hawking-temperature}) becomes
\begin{equation}\label{ss-temperature}
  T_\mathrm{BH}= \frac{\hbar}{8 \pi G k_\mathrm{B} M} \,\approx 10^{-6}
  \frac{M_\odot}{M}\ \mathrm K \,.
\end{equation}
For solar-mass black holes (and even more so for the galactic black hole),
this is of course utterly negligible -- the
black hole absorbs much more from the ubiquitous 3K-microwave background
radiation than it radiates itself.

One can, however, estimate the lifetime of a black hole by making the
plausible assumption that the decrease in mass is equal to the energy
radiated to infinity, and using Stefan-Boltzmann's law:
\[
\frac{\rmd M}{\rmd t} \propto - A T_\mathrm{BH}^4 \propto - M^2 \times
\left(\frac{1}{M} \right)^4 = -\frac{1}{M^2}\,, \]
which, when integrated, yields
\begin{equation}\label{bh-evolution}
  t(M) \propto (M_0^3-M^3) \approx M_0^3\,.
\end{equation}
Here $M_0$ is the initial mass, and it has been assumed that after the
evaporation $M \ll M_0$. Very roughly, the lifetime of a black hole is thus
given by
\begin{equation}\label{bh-lifetime}
  \tau_\mathrm{BH} \approx \left(\frac{M_0}{m_\mathrm{p}} \right)^3
  t_\mathrm{p} \approx 10^{65} \left(\frac{M_0}{M_\odot} \right)^3
  \mathrm{years}
\end{equation}
($m_\mathrm{p}$ and $t_\mathrm{p}$ denote Planck mass and Planck time:
$m_\mathrm{p} = \hbar/l_\mathrm{p}$, $t_\mathrm{p} = l_\mathrm{p}$.)
The galactic black hole thus has a lifetime of about
$3\times 10^{85}$ years! If in
the early universe primordial black holes with $M_0
\approx 5 \times 10^{14} \mathrm{g}$ were created, they would evaporate at
the present age of the universe, see \sref{primordial}.

A very detailed investigation into black-hole evaporation was made by Page
(1976). He found that for $M \gg 10^{17} \mathrm{g}$ the power emitted from
an (uncharged, non-rotating) black hole is
\[ P \approx 2.28 \times 10^{-54} L_\odot \left(\frac{M}{M_\odot}
  \right)^{-2}\,,
\]
$81.4 \%$ of which is in neutrinos (he considered only electron- and
muon-neutrinos), $16.7 \%$ in photons, and $1.9 \%$ in gravitons, assuming
that there are no other massless particles. Since a black hole
emits \textit{all} existing particles in Nature, this result would of
be changed by the existence of massless supersymmetric or other
particles. In the range $ 5 \times 10^{14} \mathrm{g} \ll M \ll 10^{17}
\mathrm{g}$, Page found
\[
  P \approx 6.3 \times 10^{16}\left(\frac{M}{10^{15} \mathrm{g}}
  \right)^{-2} \frac{\mathrm{erg}}{\mathrm{s}}\,,
\]
$45 \%$ of which is in electrons and positrons, $45 \%$ in neutrinos, $ 9
\%$ in photons, and $1 \%$ in gravitons. Massive particles with mass $m$
are only suppressed if $k_\mathrm{B} T_\mathrm{BH} < m$. For $M < 5 \times
10^{14} {\rm g}$ also higher-mass particles are emitted.

\medskip
All of the above derivations use the approximation where the spacetime
background remains classical.\footnote{This limit is referred to as the
semiclassical approximation to quantum gravity (see e.g.\ Kiefer 1994).} In
a theory of quantum gravity, however, such a picture cannot be maintained.
Since the black hole
becomes hotter while radiating, see (\ref{ss-temperature}), its mass will
eventually enter the quantum-gravity domain $ M \approx m_\mathrm{p}$,
where the semiclassical approximation breaks down. The evaporation then
enters the realm of speculation. As an
intermediate step one might consider the heuristic `semiclassical' Einstein
equations,
\begin{equation}\label{semiclassical-einstein}
  R_{ab} - \frac{1}{2} g_{ab} R = 8 \pi G \langle  T_{ab} \rangle\,,
\end{equation}
where on the right-hand side the quantum expectation value of the
energy-momentum tensor appears. The evaluation of $\langle  T_{ab} \rangle$
-- which requires regularisation and renormalisation -- is a difficult
subject on its own (Frolov and Novikov~1998). The renormalised $\langle
T_{ab} \rangle$ is essentially unique (its ambiguities can be absorbed in
coupling constants) if certain sensible requirements are imposed.
Evaluating the components of the renormalised $\langle  T_{ab} \rangle$
near the horizon, one finds that there is a flux of \textit{negative
energy} into the hole. Clearly this leads to a decrease of the black
hole's mass. These negative energies are a typical quantum effect and are
well-known from the -- accurately measured -- Casimir effect. This
occurrence of negative energies is also responsible for the breakdown of
the classical area law discussed in \sref{mechanics}.

The negative flux near the horizon lies also at the heart of the
`pictorial' representation of Hawking radiation that is often used,
see e.g. Parikh and Wilczek (2000).
In vacuum, virtual pairs of particles are created
and destroyed. However, close to the horizon, one partner of this virtual
pair might fall into the black hole, thereby liberating the other partner
to become a real particle and escaping to infinity as Hawking radiation.
The global quantum field exhibits quantum entanglement between the
in-and outside of the black hole, similar to the case of the
accelerated observer discussed above. 

\medskip
I want to end this section by giving the explicit expressions for the
Hawking temperature (\ref{hawking-temperature}) in the case of rotating and
charged black holes. For the Kerr solution, one has
\begin{equation}\label{kerr-temperature}
  k_\mathrm{B} T_\mathrm{BH} = \frac{\hbar \kappa}{2 \pi} = 2 \left( 1 +
  \frac{M}{\sqrt{M^2-a^2}} \right) ^{-1}\!\frac{\hbar}{8 \pi M} \quad< \frac{\hbar}{8 \pi
  M}\,.
\end{equation}
Rotation thus reduces the Hawking temperature. 
For the Reissner-Nordstr\"{o}m solution
(describing a charged spherically-symmetric black hole) one has
\begin{equation}\label{reissner-temperature}
    k_\mathrm{B} T_\mathrm{BH} =  \frac{\hbar}{8 \pi M}
    \left( 1 - \frac{(Gq)^4}{r_+^4} \right)  \quad< \frac{\hbar}{8 \pi
  M}\,.
\end{equation}
Thus, also electric charge reduces the Hawking temperature. For an extremal
black hole, $r_+ = GM = \sqrt G |q|$, and thus $T_\mathrm{BH} = 0$.

\section{Interpretation of entropy}
\label{entropy}

We have seen in the last
section that -- if quantum theory is taken into account -- black holes emit
thermal radiation with the temperature (\ref{hawking-temperature}).
Consequently, the laws of black-hole mechanics discussed in 
\sref{mechanics} have indeed a physical interpretation as thermodynamical
laws -- black holes \textit{are} thermodynamical systems.

 From the First Law \eref{first-law} one can therefore also infer the
expression for the black-hole entropy. From $\rmd M = T_\mathrm{BH}\rmd
S_\mathrm{BH}$ one finds the `Bekenstein-Hawking entropy'
\begin{equation}\label{bekenstein-hawking-entropy}
  S_\mathrm{BH} = \frac{k_\mathrm{B} A}{4 G \hbar}\,,
\end{equation}
in which the unknown factor in (\ref{ts-proportionalto}) has now been fixed. For the
special case of a Schwarzschild black hole, this yields
\begin{equation}\label{ss-entropy}
  S_\mathrm{BH} = \frac{k_\mathrm{B} \pi R_0^2}{ G \hbar}\,.
\end{equation}
It can easily be estimated that $S_\mathrm{BH}$ is much bigger than the
entropy of the star that collapsed to form the black hole.
The entropy of the sun, for example, is $S_{\odot}\approx
10^{57}k_B$, whereas the entropy of a solar-mass black hole
is about $10^{77}k_B$, which is twenty orders of magnitude larger!
For the galactic black hole, the entropy is 
$S_{GBH}\approx 10^{90}k_B$ which is one hundred times the entropy of the
Universe. (Under the ``entropy of the Universe'' I understand the
entropy of the present Universe up to the Hubble radius without
taking black holes into account. It is dominated by the entropy of
the cosmic microwave background radiation.)

Can a physical interpretation of this huge discrepancy be given?
Up to now the laws of black-hole
mechanics are only phenomenological thermodynamical laws. The central open
question therefore is:
 Can $S_\mathrm{BH}$ be derived from quantum-statistical
considerations? This would mean that $S_\mathrm{BH}$ could be calculated
from a Gibbs-type formula according to
\begin{equation}\label{gibbs-formula}
  S_\mathrm{BH} \overset{\mathbf{?}}{=} - k_\mathrm{B} \Tr(\rho \ln \rho) \equiv
  S_\mathrm{SM}\,,
\end{equation}
where $\rho$ denotes an appropriate density matrix; $S_\mathrm{BH}$ would
then somehow correspond to the number of quantum microstates that are
consistent with the macrostate of the black hole that is -- according to
the no-hair theorem -- uniquely characterised by mass, angular momentum,
and charge. Some important questions are:
\begin{itemize}
  \item Does $S_\mathrm{BH}$ correspond to states hidden behind the
  horizon?

  \item Or does $S_\mathrm{BH}$ correspond to the number of possible
  initial states?

  \item What are the microscopic degrees of freedom?

  \item Where are they located (if at all)?

  \item Can one understand the universality of the result?

  \item What happens with $S_\mathrm{BH}$ after the black hole has
  evaporated?

  \item Is the entropy a ``one loop'' or a ``tree level'' effect?
\end{itemize}

\medskip
The attempts to calculate $S_\mathrm{BH}$ by state counting are usually
done in the `one-loop limit' of quantum field theory in curved spacetime --
this is the limit where gravity is classical but non-gravitational fields
are fully quantum, and it is the limit where the Hawking radiation
(\ref{hawking-temperature}) has been derived. The
expression (\ref{bekenstein-hawking-entropy}) can
 already be calculated
from the so-called `tree level' of the theory, where only the gravitational
degrees of freedom are taken into account. Usually a saddle-point approximation
for a euclidean path integral is being performed. Such derivations are,
however, equivalent to derivations within classical thermodynamics,
cf. Wald (2001).

\medskip
If the entropy (\ref{bekenstein-hawking-entropy}) is to make sense, there
should be a generalised Second Law of thermodynamics according to
\begin{equation}\label{generalised-second-law}
  \frac{\rmd}{\rmd t} (S_\mathrm{BH} + S_\mathrm{M}) \ge 0 \,,
\end{equation}
where $S_\mathrm{M}$ denotes all contributions to non-gravitational
entropy. The validity of (\ref{generalised-second-law}), although far from
being proven in general, has been shown in a variety of
gedanken experiments. One of the most instructive of such experiments 
was devised by Unruh and Wald. It makes use of the box
shown in figure \ref{station} that is adiabatically lowered towards a
(spherically symmetric) black hole.

At asymptotic infinity $r \rightarrow \infty $, the black-hole radiation is
given by \eref{hawking-temperature}. However, for finite $r$ the
temperature is modified by the occurrence of a redshift factor $\chi(r)
\equiv (1-2GM/r)^{1/2}$ in the denominator. Since the box is not in free fall,
it is accelerated with an acceleration $a$. From the relation (Wald 1984)
\begin{equation}\label{kappa-limes}
  \kappa = \lim_{r \rightarrow R_0} (a \chi)\,,
\end{equation}
one has
\begin{equation}\label{temperature(radius)}
  T_\mathrm{BH}(r)= \frac{\hbar \kappa}{2 \pi k_\mathrm{B} \chi(r)} \quad
  \stackrel{r \rightarrow R_0}{\longrightarrow} \quad
  \frac{\hbar a}{2 \pi k_\mathrm{B}} \,,
\end{equation}
which is just the Unruh temperature \eref{rindler-temperature}! This means
that a freely falling observer near the horizon observes no radiation at
all, and the whole effect (\ref{temperature(radius)}) comes from the
observer (or box) being non-inertial with acceleration $a$.

The analysis of Unruh and Wald, which is a generalisation of the
gedankenexperiment discussed at the end of section \ref{mechanics}, shows
that the entropy of the black hole increases at least by the entropy of the
Unruh radiation displaced at the floating point -- this is the point where
the gravitational force (pointing downwards) and the buoyancy force from
the Unruh radiation (\ref{temperature(radius)}) are in equilibrium.
Interestingly, it is just the application of `Archimedes'principle' to this
situation that rescues the generalised Second Law
(\ref{generalised-second-law}).

 An inertial, i.e.\ free-falling, observer
does not see any Unruh radiation. How does he interpret the above result?
For him the box is accelerated and therefore the interior of the box fills
up with negative energy and pressure -- a typical quantum effect that
occurs if a `mirror' is accelerated through the vacuum.
The `floating point' is
then reached after this negative energy is so large that the total energy
of the box is zero.

\medskip
I want to conclude this section with some speculations about the final
stages of black-hole evolution and the information-loss problem. The point
is that -- in the semiclassical approximation used by Hawking -- the
radiation of a black hole seems to be
 purely thermal. If the black hole evaporates
completely and leaves only thermal radiation behind, one would have a
conflict with established principles in quantum theory: Any initial state
(in particular a pure state) would evolve into a mixed state. In ordinary
quantum theory, because of the unitary evolution of the total
system, this cannot happen. Formally, $\Tr\rho^2$ remains \textit{constant}
under the von Neumann equation; the same is true for the entropy
$S_\mathrm{SM} = -k_\mathrm{B}\Tr(\rho\ln\rho)$: For a unitarily evolving
system, there is no increase in entropy. If these laws are violated during
black-hole evaporation, information would be destroyed. This is indeed the
speculation that Hawking made after his discovery of black-hole radiation.
The attitudes towards this \textit{information-loss problem} can be roughly
divided into the following classes,
\begin{itemize}
  \item The information is indeed lost during black-hole evaporation, and
  the quantum-mechanical Liouville equation is replaced by an equation of
  the form
\begin{equation}\label{new-liouville}
  \rho \longrightarrow S\mspace{-15mu}\parallel \rho \neq S \rho S^\dagger\,.
\end{equation}

  \item The full evolution is in fact unitary; the black-hole radiation
  contains subtle quantum correlations that cannot be seen in the
  semiclassical approximation.

  \item The black hole does not evaporate completely, but leaves a
  `remnant' with mass in the order of the Planck mass that carries the
  whole information.
\end{itemize}

In my opinion, the information-loss problem is only a
pseudoproblem. Already in the original calculation of Hawking (1975)
only pure states appear. Reference to thermal radiation is being made
because the particle number operator in the final pure state possesses
an exact Planckian distribution. As has been shown in Kiefer (2001) 
the coupling of this pure state (a squeezed state in quantum-optics
language) to its natural environment produces a thermal ensemble
for the Hawking radiation, which constitutes an open quantum system,
 after this environment is traced out.
The thermal nature of this radiation is thus a consequence of
decoherence (Giulini \etal~1996).

There exist many attempts to derive the Bekenstein-Hawking entropy
\eref{bekenstein-hawking-entropy} within approaches to quantum gravity,
see e.g. Kiefer (1999) and Wald (2001) for more details and references.
Examples are the derivations within superstring theory (counting
of states referring to microscopic objects called D-branes),
canonical quantum gravity, Sakharov's induced gravity, conformal field
theories, and others. Although many of these look very promising,
a final consensus has not yet been reached.
 
\section{Primordial black holes}
\label{primordial}

Can the above discussed quantum effects of black holes be observed?
As has already been mentioned, black holes formed by stellar collapse
are much too heavy to exhibit quantum behaviour. To form smaller black holes
one needs higher densities which can only occur under the extreme
situations of the early Universe\footnote{In theories with
large extra dimensions it is imaginable that quantum effects
of black holes can be seen at ordinary accelerators, see
Dimopoulos and Landsberg~2001.}.
Such primordial black holes can originate in the radiation-dominated
phase during which no stars or other objects can be formed. 

\begin{figure}[!t]
 \begin{center}
  \psfrag{yachse}[][][0.8]
         {\begin{tabular}{c}$\log$\\$\lambda_{phys}$\\ \end{tabular}}
  \psfrag{xachse}{$\log a(t)$}
  \psfrag{ai}[][][0.8]{$a_i$}
  \psfrag{af}[][lB][0.8]{$a_f$}
  \psfrag{tki}[][lt][0.65]{$t_{k,exit}$}
  \psfrag{tkf}[][rt][0.65]{$t_{k,enter}$}
  \psfrag{aeq}[][cB][0.8]{$a_{eq}$}
  \psfrag{lambda}[lc][][0.7]{$\lambda_k(t):=\frac{a(t)}{k}$}
  \psfrag{H-1}[][][0.8]{$H^{-1}(t)$}
  \framebox[0.7\textwidth]{\includegraphics[width=0.62\textwidth]
  {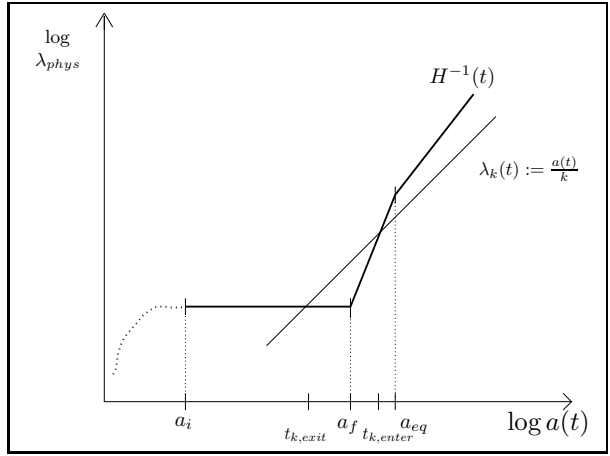}}
 \end{center}
 \caption[]{Time development of a physical scale
 $\lambda(t)$ and the Hubble horizon $H^{-1}(t)$. During an
 inflationary phase $H^{-1}(t)$ remains approximately constant.
 After the end of inflation
 ($a_f$) the horizon $H^{-1}(t)$ increases faster than any scale.
 Therefore $\lambda_k$ enters the horizon again at $t_{k,enter}$
 in the radiation- (or matter-) dominated phase.
 \label{skalen}} 
\end{figure}

\medskip
Consider for simplicity a spherically-symmetric region with radius
$R$ and density $\rho=\rho_c+\delta\rho$ embedded in a flat Universe with the
critical density $\rho_c$, cf. Carr (1985).
 For spherical symmetry the inner region
is not affected by matter in the surrounding part of the Universe,
so it will behave like a closed Friedmann Universe (since its density
is overcritical), i.e., the expansion of this region will come to a
halt at some stage, followed by a collapse. In order to reach a
complete collapse, the (absolute value of the)
potential energy, $V$, at the time of maximal
expansion has to exceed the inner energy, $U$, given by the pressure $p$.
I.e.,
\begin{equation}
V \sim \frac{GM^2}{R} \sim G\rho^2R^5 \gtrsim pR^3\ .
\end{equation}
If the equation of state reads $p=wR$ ($w=1/3$ for radiation dominance), this
gives
\begin{equation}
R\gtrsim \sqrt{w}\frac{1}{\sqrt{G\rho}} \ .
\end{equation}  
The lower bound for $R$ is thus just given by the {\em Jeans length}.
There also exists an upper bound. The reason is that $R$ must be smaller
than the curvature radius (given by $1/\sqrt{G\rho}$)
of the overdense region at the moment of collapse. Otherwise the region
would contain a compact three-sphere which is topologically disconnected
from the rest of the Universe. This case would then not lead to a black hole
within our Universe. Using $\rho\sim\rho_c\sim H^2/G$, where $H$ denotes
the Hubble parameter of the background flat Universe, one has
the condition
\begin{equation}
H^{-1}\gtrsim R\gtrsim \sqrt{w}H^{-1}\ ,
\label{bounds}
\end{equation}
evaluated at the time of collapse, for the formation of a black hole.
This relation can be rewritten also as a condition referring to any
initial time of interest (Carr~1985). In particular, one is often interested
in the time where the fluctuation enters the horizon in the
radiation-dominated Universe. This is illustrated in figure~\ref{skalen},
where the presence of a possible inflationary phase at earlier times
is also shown.

At horizon entry one gets, denoting $\delta\equiv\delta\rho/\rho_c$,
\begin{equation}
1\gtrsim \delta_{enter} \gtrsim 0.3 \ .
\end{equation}
This is, however, only a rough estimate. Numerical calculations give
instead the bigger value $\delta_{min}\approx 0.7$ (Niemeyer and
Jedamzik~1999). 

Taking from (\ref{bounds}) $R\approx \sqrt{w}H^{-1}$, one gets for the
initial mass of a primordial black hole (PBH),
\begin{equation}
M_{PBH}= \frac{4\pi}{3}\rho R^3 \approx \frac{4\pi}{3}\rho_c(1+\delta)
w^{3/2}H^{-3} \approx w^{3/2}M_H\ ,
\label{MPBH}
\end{equation}  
where $M_H\equiv (4\pi/3)\rho_cH^{-3}$ denotes the mass inside the horizon.
Since $M_{PBH}$ is of the order of this horizon mass, a collapsing region
will form a black hole practically immediately after horizon entry.
Using the relation $M_H=t/G$, valid for a radiation dominated Universe,
one gets from (\ref{MPBH}) the quantitative estimate 
\begin{equation}
M_{PBH}[{\rm g}]\approx 10^{38}\ t[{\rm s}]\ .
\end{equation}
This means that one can create Planck-mass black holes at the
Planck time, and PBHs with $M_{PBH}\approx 5\times 10^{14}{\rm g}$
at $t\approx 5\times 10^{-24}{\rm s}$. The latter value is important,
since according to \eref{bh-lifetime} black holes with masses smaller than
  $M_{PBH}\approx 5\times 10^{14}{\rm g}$ have by now evaporated due to 
Hawking radiation. PBHs with bigger mass are still present today.
At $t\approx 10^{-5}{\rm s}$ one can create a solar-mass black hole
and at $t\approx 10{\rm s}$ (the time of nucleosynthesis) one could
form a PBH with the mass of the galactic black hole. 
The initial mass can of course increase through accretion,
but it turns out that this is negligible in most circumstances
(Carr~1985). 

In the presence of an inflationary phase in the early Universe,
all PBHs produced before the end of inflation are diluted away. This gives
the bound
\begin{equation}
M_{PBH} > M_H(T_{RH}) \approx \frac{m_p^3}{10.88 T_{RH}^2}
\sim 1{\rm g}\ ,
\end{equation}
if for the reheating temperature $T_{RH}$ a value of 
$10^{16}{\rm GeV}$ is chosen.  

According to the numerical calculations by Niemeyer and Jedamzik (1999),
there exists a whole spectrum of initial masses,
\begin{equation}
M_{PBH}=KM_H(\delta-\delta_{min})^{\gamma}\ ,
\end{equation}
a relation that is reminiscent of the theory of critical phenomena.
This may change some of the quantitative conclusions.

\begin{figure}[t]
 \begin{center}
  \psfrag{y-Achse}[][][0.7]{$\lg \alpha_{max}$}
  \psfrag{x-Achse}[][][0.7]{$\lg M \mathrm{[g]}$}
  \framebox[0.7\textwidth]{\includegraphics[width=0.65\textwidth]
    {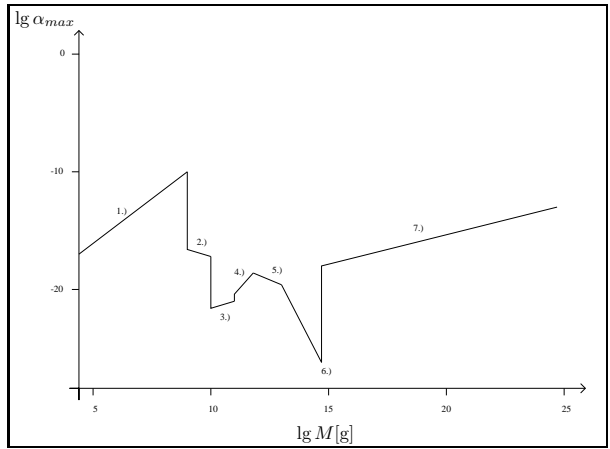}}
 \end{center}
 \caption[]{Strongest constraints on the initial PBH mass fraction.
  The numbers correspond to the various entries in
  table \ref{tpbhconstraints}.
 \label{abbpbhconstraints}} 
\end{figure}

To calculate the production rate of PBHs one needs an initial spectrum
of fluctuations. This is usually taken to be of a Gaussian form,
as predicted by most inflationary models (cf. Liddle and Lyth~2000).
Therefore, there exists always a nonvanishing probability that
the density contrast is high enough to form a black hole,
even if the maximum of the Gaussian corresponds to a small value.
 One can then calculate the mass ratio (compared to the total mass)
of regions which will develop into PBHs with mass $M_{PBH}\gtrsim
M$, see, e.g., Bringmann \etal (2001), Sec.~2, for details. This mass ratio,
given by
\begin{equation}
\alpha(M):=\frac{\rho_{PBH,M}}{\rho_r}\approx\Omega_{PBH,M}
                \equiv\frac{\rho_{PBH,M}}{\rho_c}\ ,
\end{equation}
where $\rho_r$ is the radiation density, is then compared with
observation. This, in turn, gives a constraint on the theoretically
calculated initial spectrum. 
Table \ref{tpbhconstraints} presents
various observational constraints on $\alpha$ (see Green and Liddle~1997).
The corresponding maximal value for each $\alpha$ is shown for the various
constraints in figure~\ref{abbpbhconstraints}.

\begin{table}[t]
  \begin{center}
  \setlength{\tabcolsep}{1.8ex}
  \renewcommand{\arraystretch}{1.2}
  \begin{tabular}{l r @{$M$} l l}
    \hline \hline
      \multicolumn{1}{c}{constraint}& 
      \multicolumn{2}{c}{range} &
      \multicolumn{1}{c}{reason}\\
    \hline
    $\alpha<0.1(M/10^{15}\mbox{ g})^\frac{3}{2}$ & & $<10^{15}$ g
            & radiation relics \\
    $\alpha<10^{-17}(10^9\mbox{ g}/M)^\frac{1}{2}$
            & $10^9$ g$<$ & $<10^{11}$ g & $n_n/n_p$-ratio\\
    $\alpha<10^{-22}(M/10^{10}\mbox{ g})^\frac{1}{2}$
            & $10^{10}$ g$<$ & $<10^{11}$ g & deuterium dissociation\\
    $\alpha<10^{-21}(M/10^{11}\mbox{ g})^\frac{5}{2}$
            & $10^{11}$ g$<$ & $<10^{13}$ g & helium fission\\
    $\alpha<10^{-16}(10^9\mbox{ g}/M)$& $10^9$ g$<$ & $<10^{13}$ g
            & entropy per baryon\\
    $\alpha<10^{-26}$& \multicolumn{2}{c}{$M\approx5\times 10^{14}$ g}
            & $\gamma$ background\\
    $\alpha<10^{-18}(M/10^{15}\mbox{ g})^\frac{1}{2}$ & & $>10^{15}$ g
            & present PBH density \\
        \hline \hline  
  \end{tabular} 
  \end{center}
 \caption{Constraints on the mas fraction
          $\alpha(M):=\frac{\rho_{PBH,M}}{\rho_r}\approx \Omega_{PBH,M}$  
          of primordial black holes at their time of formation
          (Green and Liddle~1997).
\label{tpbhconstraints}} 
\end{table}

Constraints arise either from Hawking radiation or from the gravitational
contribution of PBHs to the present Universe (last entry).
PBHs with initial mass of about $5\times 10^{15}{\rm g}$ evaporate ``today''.
(They release about $10^{30}{\rm erg}$ in the last second.)
 From observations of the $\gamma$-ray background one can find the
constraint given in the table. It corresponds to an upper limit
of about $10^4$ PBHs per kubic parsec or $\Omega_{PBH,0}<10^{-8}$.
One can also try to observe directly the final evaporation event of
a single PBH. This gives an upper limit of about 
$4.4\times 10^5$ events per kubic parsec per year.

Given these observational constraints, one can then calculate
the ensuing constraints on the primordial spectrum. The gravitational
constraint $\Omega_{PBH,0}<1$ gives surprisingly strong
restrictions (cf. Bringmann \etal~2001). For a scale-free spectrum
of the form $\propto k^n$, as it is usually discussed for inflationary
models, one finds restrictions on $n$ that are comparable to
the limits obtained by large-scale observations (anisotropy spectrum
of the cosmic microwave background radiation). Since these restrictions
come from observational constraints referring to much smaller scales,
they constitute an important complementary test. 

The question whether PBHs really exist in nature has thus not yet been
settled. Their presence would be of an importance that could hardly be
overestimated. They would give the unique opportunity to study the
quantum effects of black holes and could yield the crucial key for
the construction of a final theory of quantum gravity.

\section*{References}

\begin{thereferences}

\item Bekenstein J D 1980 {\em Physics Today} (January) 24

\item Bringmann T, Kiefer C and Polarski D 2001
      {\em Phys. Rev.~D} {\bf 65} 024008

\item Carr B J 1985 {\em Observational and theoretical aspects
      of relativistic astrophysics and cosmology} ed J L Ganz and
      L J Goicoechea (Singapore: World Scientific) p~1

\item Chen P and Tajima T 1999 {\em Phys. Rev. Lett.} {\bf 83} 256

\item Davies P C W 1975 {\em J. Phys.~A} {\bf 8} 609

\item Dimopoulos S and Landsberg G 2001
        {\em Phys. Rev. Lett.} {\bf 87} 161602

\item Farrugia Ch J and Hajicek P 1979 {\em Commun. Math. Phys.}
      {\bf 68} 291

\item Frolov V P and Novikov I D 1998 {\em Black hole physics}
      (Dordrecht: Kluwer)

\item Fulling S A 1973 {\em Phys. Rev.~D} {\bf 7} 2850

\item Giulini D, Joos E, Kiefer C, Kupsch J, Stamatescu I O and
      Zeh H D 1996 {\em Decoherence and the appearance of a classical
      world in quantum theory} (Berlin: Springer) 

\item Green A M and Liddle A R 1997 {\em Phys. Rev. D} {\bf 56} 6166

\item Hawking S W 1975 {\em Commun. Math. Phys.} {\bf 43} 199

\item Hehl F W, Kiefer C and Metzler R (eds) 1998 {\em Black Holes:
      Theory and observation} (Berlin: Springer)

\item Heusler M 1996 {\em Black hole uniqueness theorems}
      (Cambridge: Cambridge University Press)

\item Kiefer C 1994 {\em Canonical gravity: From classical to quantum}
      ed J Ehlers and H Friedrich (Berlin: Springer) p~170

\item Kiefer C 1999 {\em Classical and quantum black holes} 
      ed P Fr\'e \etal (Bristol: IOP) p~18

\item Kiefer C 2001 {\em Class.
      Quantum Grav.} {\bf 18} L151

\item Leinaas J M 2001 {\tt hep-th/0101054}

\item Liddle A R and Lyth D H 2000 {\em Cosmological inflation and
      large-scale structure} (Cambridge: Cambridge University Press)

\item Niemeyer J C and Jedamzik K 1999 {\em Phys. Rev. D} {\bf 59}
      124013

\item N\'{u}\~{n}ez D, Quevedo H and Sudarsky D 1998 {\em in}
      Hehl \etal (1998) p~187

\item Page D N 1976 {\em Phys. Rev. D} {\bf 13} 198

\item Parikh M K and Wilczek F 2000 {\em Phys. Rev. Lett.} {\bf 85} 5042

\item Sexl R U and Urbantke H K 1983 {\em Gravitation und Kosmologie}
      (Mannheim: Bibliographisches Institut)

\item Thorne K S 1994 {\em Black holes and time warps} (New York: Norton)

\item Unruh W G 1976 {\em Phys. Rev.~D} {\bf 14} 870

\item Wald R M 1984 {\em General Relativity} (Chicago: University of
      Chicago Press)

\item Wald R M 2001 {\em The thermodynamics of black holes}\\
      http://www.livingreviews.org/Articles/Volume4/2001-6wald

\end{thereferences}

\end{document}